\documentclass[aip,apl,reprint,twoside,floatfix,showpacs,superscriptaddress]{revtex4-1}

\usepackage[english]{babel}
\usepackage{graphicx}
\usepackage{amsmath}
\usepackage{amssymb}
\usepackage{amsfonts}
\usepackage{dcolumn}
\usepackage{epsfig}
\usepackage{subfigure}
\usepackage{bm}
\usepackage[version=3]{mhchem}
\usepackage{textcomp}
\usepackage{miller}
\usepackage{gensymb}
\usepackage{array}

\newcommand{\Wt}{\ensuremath{W_\mathrm{t}}}
\newcommand{\Wopt}{\ensuremath{W_\mathrm{opt}}}
\newcommand{\Wph}{\ensuremath{W_\mathrm{ph}}}

\begin{document}

\title{Charge transport in amorphous \ce{Hf_{0.5}Zr_{0.5}O2}}

\author{D.~R.~Islamov}\email{damir@isp.nsc.ru}
	\affiliation{Rzhanov Institute of Semiconductor Physics,
		Siberian Branch of Russian Academy of Sciences,
		Novosibirsk, 630090, Russian Federation}%
	\affiliation{Novosibirsk State University,
		Novosibirsk, 630090, Russian Federation}%
\author{T.~V.~Perevalov}
	\affiliation{Rzhanov Institute of Semiconductor Physics,
		Siberian Branch of Russian Academy of Sciences,
		Novosibirsk, 630090, Russian Federation}%
	\affiliation{Novosibirsk State University,
		Novosibirsk, 630090, Russian Federation}
\author{V.~A.~Gritsenko}\email{grits@isp.nsc.ru}
	\affiliation{Rzhanov Institute of Semiconductor Physics,
		Siberian Branch of Russian Academy of Sciences,
		Novosibirsk, 630090, Russian Federation}%
	\affiliation{Novosibirsk State University,
		Novosibirsk, 630090, Russian Federation}%
\author{C.~H.~Cheng}
	\affiliation{Dept. of Mechatronic Engineering, National Taiwan Normal University, Taipei, 106, Taiwan ROC}%
\author{A.~Chin}\email{albert\_achin@hotmail.com}
	\affiliation{National Chiao Tung University, Hsinchu, 300, Taiwan ROC}%

\date{\today}

\begin{abstract}
In this study, we demonstrated experimentally and theoretically
that the charge transport mechanism in amorphous \ce{Hf_{0.5}Zr_{0.5}O2} is
phonon-assisted tunneling between traps
like in \ce{HfO2} and \ce{ZrO2}.
The thermal trap energy of $1.25$\,eV
and optical trap energy of $2.5$\,eV
in
\ce{Hf_{0.5}Zr_{0.5}O2} were
determined based on 
comparison of experimental data on transport
with different theories of charge transfer in dielectrics.
A hypothesis that
oxygen vacancies are responsible for the
charge transport in \ce{Hf_{0.5}Zr_{0.5}O2}
was discussed.
\end{abstract}



\maketitle

Knowledge about charge transport of
high-$\kappa$ dielectrics is very important for
modern microelectronics. Previous transport studies
were based on binary compound like \ce{HfO2} and \ce{ZrO2}
\cite{NovikovHfO2:JAP113:024109, HfO2:Tranport:2014, ZrO2:Transport:jap109:014504}.
It was shown, that charge transport mechanism
in \ce{HfO2} and \ce{ZrO2}
is phonon-assisted tunneling between
traps \cite{HfO2:Tranport:2014, ZrO2:transport:2015}.
The charge transport mechanism of ternary high-$\kappa$
solid solution
\ce{Hf_{0.5}Zr_{0.5}O2} still remains unknown.

In this letter, we investigate the charge transport
mechanism in \ce{Hf_{0.5}Zr_{0.5}O2} by comparison of experimental data
with different theories of charge transfer in dielectrics.

Transport measurements were performed for structures
\ce{Si}/\ce{Hf_{0.5}Zr_{0.5}O2}/\ce{Ni}.
To fabricate these structures, we deposited
the 20-nm-thick \ce{Hf_{0.5}Zr_{0.5}O2} solid solution
films
on  $n$- and $p$-type \ce{Si} \hkl(100) wafers
by physical vapor deposition (PVD).
Pure \ce{HfO2} and \ce{ZrO2} targets were bombarded
by electron beams in high vacuum chamber,
and \ce{Hf_{x}Zr_{y}O} were deposited on the wafer
forming \ce{Hf_{0.5}Zr_{0.5}O2} films.
A \ce{Zr}/\ce{Hf} ratio of $\simeq 1$ was used.
We did not apply any post-deposition annealing
to produce the most non-stoichiometric films.
The structural properties of grown high-$\kappa$
\ce{Hf_{0.5}Zr_{0.5}O2}
dielectric were examined by grazing incidence x-ray
diffraction diffractogram (GI-XRD).
The structural analysis showed that the resulting
\ce{Hf_{0.5}Zr_{0.5}O2} films were amorphous.
All samples for transport measurements were equipped with round
$50$-nm-thick Ni gates with a radius of $70$\,$\mu$m.
The measurements were performed using a Hewlett Packard 4155B
semiconductor parameter analyzer and an Agilent E4980A precision LCR meter.

Optical (dynamic) permittivity $\varepsilon_\infty$
of \ce{Hf_{0.5}Zr_{0.5}O2} was
calculated in the framework of density functional theory
using the \textsl{ab initio} simulation code 
{\sc Quantum ESPRESSO} \cite{QE-2009}.
Electronic structures
and dielectric properties 
of monoclinic \ce{Hf_{0.5}Zr_{0.5}O2} 
using $12$-atom cell was simulated. 
The structure was obtained by replacement
of a half hafnium atom to zirconium
in monoclinic primitive cell
of \ce{HfO2} with following relaxation.
The similar approach was described earlier \cite{ZrO2:1principle:epsilon:jap105:106103}.

\begin{figure}
  \includegraphics[width=\columnwidth]{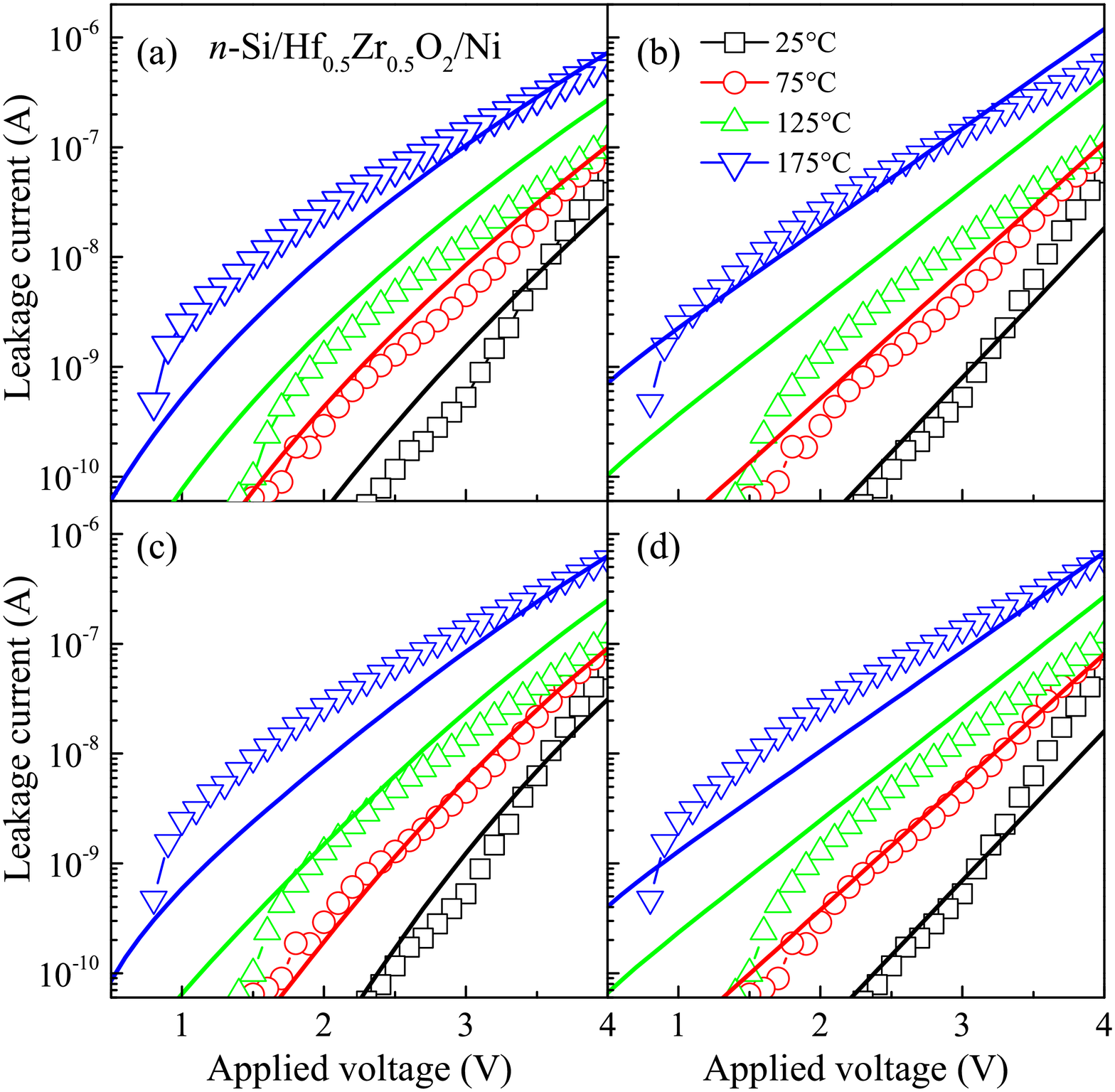}
  \subfigure{\label{f:IVT:n-Si_HfZrO_Ni:Frenkel}}
  \subfigure{\label{f:IVT:n-Si_HfZrO_Ni:Hill}}
  \subfigure{\label{f:IVT:n-Si_HfZrO_Ni:Makram}}
  \subfigure{\label{f:IVT:n-Si_HfZrO_Ni:Nasyrov}}
  \caption{Experimental (characters) and
    simulations (solid lines)
    cur\-rent-vol\-tage characteristics
    in $n$-\ce{Si}/\ce{Hf_{0.5}Zr_{0.5}O2}/\ce{Ni} structures
    at different temperatures.
    \subref{f:IVT:n-Si_HfZrO_Ni:Frenkel} Frenkel model (\ref{eq:P:Frenkel}), 
    \subref{f:IVT:n-Si_HfZrO_Ni:Hill} Hill model of overlapped traps (\ref{eq:P:Hill}),
    \subref{f:IVT:n-Si_HfZrO_Ni:Makram} multiphonon trap ionization (\ref{e:P:Makram}),
    \subref{f:IVT:n-Si_HfZrO_Ni:Nasyrov} phonon-assisted tunneling between traps (\ref{eq:P:Nasyrov}).
  }
  \label{f:IVT:n-Si_HfZrO_Ni}
\end{figure}

Fig.~\ref{f:IVT:n-Si_HfZrO_Ni:Frenkel} shows a set of experimental
current-voltage characteristics ($I$-$V$) of
$n$-\ce{Si}/\ce{Hf_{0.5}Zr_{0.5}O2}/\ce{Ni} structures
measured at different temperatures $T$
by characters in various shapes and colors.
Positive applied voltage corresponds to positive bias on the
\ce{Ni} contact.
The leakage current through \ce{Hf_{0.5}Zr_{0.5}O2} grows exponentially
with increasing of electric field (or applied voltage)
and temperature in accumulation mode ($V>0$).

Experiment results were analyzed
by using different models of charge transport in
dielectrics:
\begin{equation}
  I = eSN^{2/3}P,
  \label{eq:I}
\end{equation}
where
$I$ is the full current through the sample,
$e$ is the elementary charge,
$S=\pi(70\,\mu\text{m})^2$ is the contact square,
$N$ is the bulk trap density,
and $P$ is the probability rate of charge carrier transfer
between traps, which depends on the transport model.
A mathematical model of well known Frenkel law was
introduced in 1938
for isolated trap ionization \cite{PhysRev:54:647}:
\begin{equation}
  P = \nu\exp\left(\frac{W-\beta_\mathrm{F}\sqrt{F}}{kT} \right), \beta_\mathrm{F} = \frac{e^3}{\pi\varepsilon_0\varepsilon_\infty},
  \label{eq:P:Frenkel}
\end{equation}
where
$\nu$ is the frequency factor which was defined as
$\nu\simeq W/h$,
$W$ is thermal ionization energy of the trap,
$h=2\pi\hbar$ is the Planck constant,
$\beta_\mathrm{F}$ is Frenkel coefficient,
$F=V/d$ is the electric field,
$d$ is the dielectric film thickness,
$k$ is the Boltzmann constant,
$\varepsilon_\infty$ is dynamic permittivity
of the dielectric film,
and $\varepsilon_0$ is vacuum permittivity (electric constant).
Results of simulations (\ref{eq:I})+(\ref{eq:P:Frenkel})
are shown in Fig.~\ref{f:IVT:n-Si_HfZrO_Ni:Frenkel}
by solid lines.
One can see, that Frenkel model
describes the experiment data
qualitatively very good.
However, quantitative fitting procedure
returns underestimated fitting parameter values:
the slopes of the fitting lines with Frenkel coefficient
give the dynamic permittivity $\varepsilon_\infty=3.6\div 3.7$,
which is lower than
$\varepsilon_\infty(\ce{HfO2})=4.4$ \cite{high-k:Samares:2013}, 
$\varepsilon_\infty(\ce{ZrO2})=5.6$ \cite{ZrO2:abinitio:PRB64:134301}, 
and calculated from the first principals
$\varepsilon_\infty(\ce{Hf_{0.5}Zr_{0.5}O2})=4.8\div 5.2$.
Further fittings return
$N=10^{7}$\,cm$^{-3}$ and $W=0.8$\,eV.
Found values the charge trap density of
$N=10^{7}$\,cm$^{-3}$
at $\nu\sim W/h\simeq 2\times 10^{14}$\,s$^{-1}$
corresponds to mean distance between traps
$s=N^{-1/3} \simeq 50$\,$\mu$m
that is comparable to \ce{Ni} gate size.
Taking all these facts into account
one can conclude that 
there is no quantitative agreement between experiments
and Frenkel model,
despite that Frenkel model
describes the experiment data qualitatively.

Simulating in terms 
of overlapped traps ionization (Hill model) \cite{PhilMag:23:59}
\begin{equation}
  P = \nu\exp\left(-\frac{W-e^2/\pi\varepsilon_0 \varepsilon_\infty s}{kT} \right)
   2 \sinh\left( \frac{esF}{2kT} \right),
  \label{eq:P:Hill}
\end{equation}
are in
good quantitative agreement with experiments
as well as Frenkel model
(Fig.~\ref{f:IVT:n-Si_HfZrO_Ni:Hill}).
However, too low value of frequency
factor of $\nu\sim 10^{7}$\,s$^{-1}$
was obtained. All values of obtained filling parameters
are collected in Table~\ref{t:values}.

Results of simulations by
the model of multiphonon trap ionization \cite{PhysRevB:25:6406}
\begin{equation}
 \begin{array}{c}
  P=  \sum\limits_{n=-\infty}^{+\infty} \exp\left( \frac{n\Wph}{2kT} - \frac{\Wopt-\Wt}{\Wph}\coth\frac{\Wph}{2kT} \right)\times\\
     \times I_n\left( \frac{(\Wopt-\Wt)/\Wph}{\sinh(\Wph/2kT)} \right) P_i(\Wt+n\Wph), \\
  P_i(W) = \frac{eF}{2\sqrt{2m^* \cdot W}} \exp \left(-\frac{3}{4} \frac{\sqrt{2m^*}}{\hbar eF} W^{3/2} \right),
 \end{array}
 \label{e:P:Makram}
\end{equation}
are shown in Fig.~\ref{f:IVT:n-Si_HfZrO_Ni:Makram}
by solid lines.
Here $\Wph$ is phonon energy,
$\Wopt$ is optical energy of the trap,
$\Wt$ is thermal trap energy,
$I_n$ are modified Bessel functions,
$m^*$ is the effective mass,
and $P_i(W)$ is probability of tunneling trough
a triangle barrier of $W$ height.
Calculated set of $I$-$V$-$T$ curves
is very close to experimental data,
obtained values of fitting parameters
include low trap density of
$N = 2\times 10^{13}$\,cm$^{-3}$
which corresponds to
$s=370$\,nm. This mean distance
between traps is much greater than
the film thickness of $20$\,nm
(Table~\ref{t:values}).
Thus, it can be concluded that
multiphonon trap ionization
does not
adequately describe charge transfer in \ce{Hf_{0.5}Zr_{0.5}O2}.

\begin{figure}
  \includegraphics[width=0.5\columnwidth]{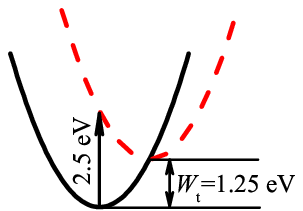}
  \caption{Configuration coordination energy diagram of trap ionization process
   on negative charged trap in \ce{Hf_{0.5}Zr_{0.5}O2}.
   Lower term is filled ground state,
   upper term is excited empty state.}
  \label{f:configdiag:thermal}
\end{figure}

To get complete vision on the charge transport
in \ce{Hf_{0.5}Zr_{0.5}O2} experiment data was simulated
basing on phonon-assisted
tunneling between traps \cite{JAP:109:093705}:
\begin{equation}
\begin{aligned}
  P = & \frac{\sqrt{2\pi}\hbar \Wt}{m^* s^2 \sqrt{\Wopt-\Wt}}
  \exp\left(-\frac{\Wopt-\Wt}{2kT} \right)\times\\
  \times & \exp\left(-\frac{2s\sqrt{m ^* \Wt}}{\hbar} \right)
  \sinh\left(\frac{eFs}{2kT} \right),
  \label{eq:P:Nasyrov}
  \end{aligned}
\end{equation}
Results of this procedure are shown in
Fig.~\ref{f:IVT:n-Si_HfZrO_Ni:Nasyrov}.
The experiment data were described quantitatively
and qualitatively with
the following values of fitting parameters (Table~\ref{t:values}):
$N = 3\times 10^{19}$\,cm$^{-3}$,
$\Wt=1.25$\,eV,
$\Wopt=2.5$\,eV,
$m^*/m_\mathrm{e}=0.23$
($m_\mathrm{e}$ is a free electron mass).
Fig.~\ref{f:configdiag:thermal}
shows the configuration diagram of
a negatively charged electron trap.
A vertical transition with a value of $2.5$\,eV corresponds
to the optical trap excitation,
transitions of $1.25$\,eV correspond to thermal trap energy.

\begin{figure}
  \includegraphics[width=\columnwidth]{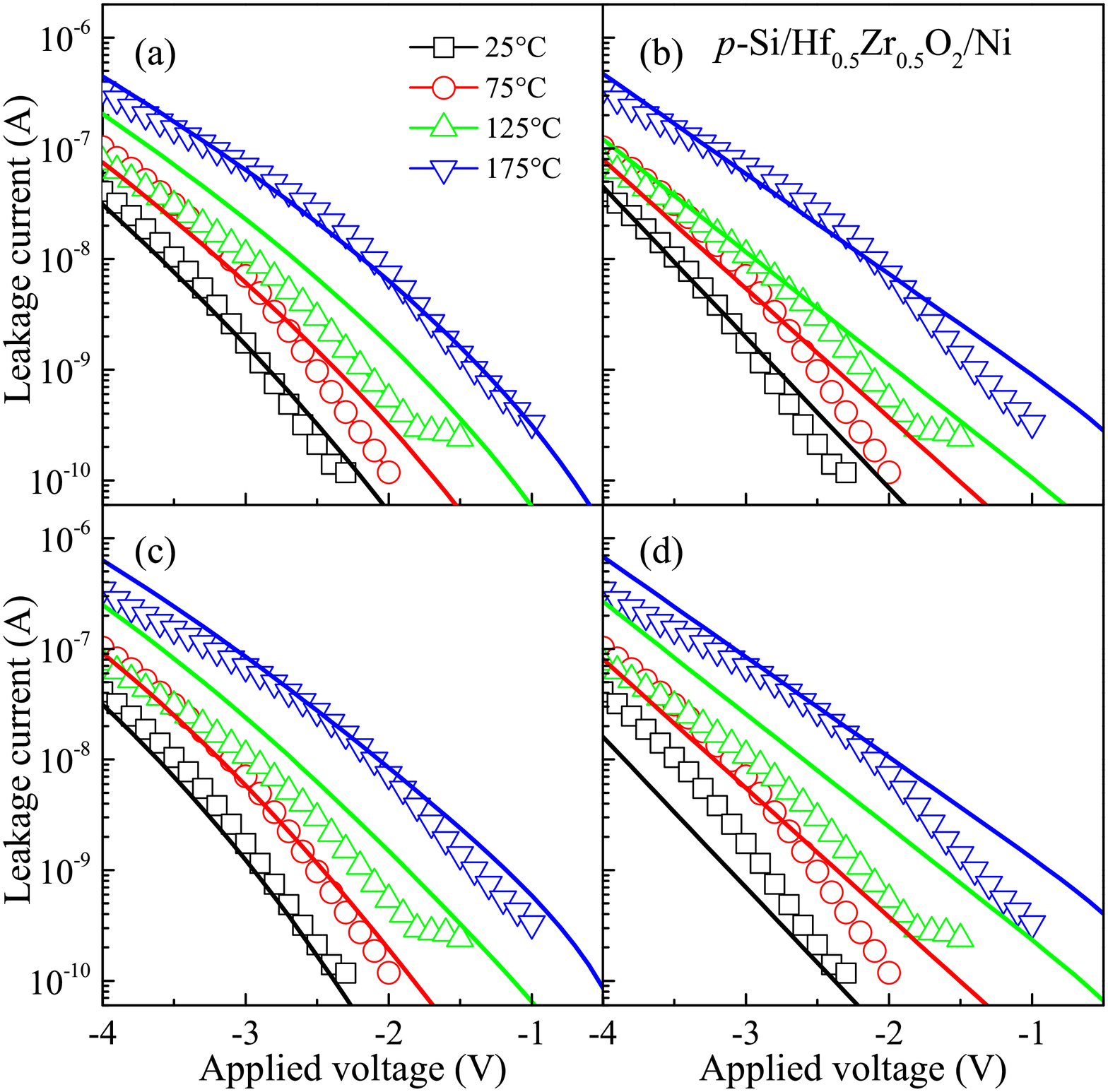}
  \subfigure{\label{f:IVT:p-Si_ZrO2_Ni:Frenkel}}
  \subfigure{\label{f:IVT:p-Si_ZrO2_Ni:Hill}}
  \subfigure{\label{f:IVT:p-Si_ZrO2_Ni:Makram}}
  \subfigure{\label{f:IVT:p-Si_ZrO2_Nim:Nasyrov}} 
  \caption{Experimental (characters)
    cur\-rent-vol\-tage characteristics
    and simulations (lines)
    in $p$-\ce{Si}/\ce{Hf_{0.5}Zr_{0.5}O2}/\ce{Ni} structures at different temperatures.
    \subref{f:IVT:n-Si_HfZrO_Ni:Frenkel} Frenkel model (\ref{eq:P:Frenkel}), 
    \subref{f:IVT:n-Si_HfZrO_Ni:Hill} Hill model of overlapped traps (\ref{eq:P:Hill}),
    \subref{f:IVT:n-Si_HfZrO_Ni:Makram} multiphonon trap ionization (\ref{e:P:Makram}),
    \subref{f:IVT:n-Si_HfZrO_Ni:Nasyrov} phonon-assisted tunneling between traps (\ref{eq:P:Nasyrov}).}
  \label{f:IVT:p-Si_ZrO2_Ni}
\end{figure}

The same measurements and simulations
were performed for
$p$-\ce{Si}/\ce{Hf_{0.5}Zr_{0.5}O2}/\ce{Ni} structures.
Results are represented in Fig.~\ref{f:IVT:p-Si_ZrO2_Ni}.
One can see that all model describe experimental curves
qualitatively.
Calculated values of fitting parameters are summarized in
Table~\ref{t:values}.
Models of isolated (\ref{eq:P:Frenkel})
and overlapped (\ref{eq:P:Hill}) charged traps
can describe experiments with
inadequate parameters like for 
$n$-\ce{Si}-based samples.
Multiphonon ionization of neutral trap (\ref{e:P:Makram})
has good agreement with
experiments at
$N = 2\times 10^{13}$\,cm$^{-3}$,
which is equal than one got for 
$n$-\ce{Si}/\ce{Hf_{0.5}Zr_{0.5}O2}/\ce{Ni} structure
by the same transport model.
At the same time calculated curves in terms of
phonon-assisted tunneling between traps (\ref{eq:P:Nasyrov})
are close to experimental data with
the same parameter values as that obtained
for $n$-\ce{Si}/\ce{Hf_{0.5}Zr_{0.5}O2}/\ce{Ni} samples.

\begin{table*}
  \centering
  \begin{tabular}{c|cccccccc}\hline\hline
  Model \vphantom{\Large{N}}&$N$ (cm$^{-3})$&$s$&$W$ (eV)&\Wt (eV)&\Wopt (eV)& $\nu$ (s$^{-1}$)&$\varepsilon_\infty$&$m^*/m_\mathrm{e}$ \\\hline
  F \vphantom{\Large{N}}&$1\times 10^{7}$&$50$\,$\mu$m&$0.8$&---&---&$\sim10^{14}$&$3.6\div 3.7$&---\\ 
  Hill &$3\times 10^{19}$&$3.2$\,nm&$0.9\div 1.0$&---&---&$\sim 10^{6}\div 10^{7}$&$5$&---\\ 
  MPTI &$2\times 10^{13}$&$370$\,nm&---&$0.8$&$1.6$&---&---&$0.17$ \\
  PAT &$3\times 10^{19}$&$3.2$\,nm&---&$1.25$&$2.5$&---&---&$0.23$ \\ \hline
  Exp \vphantom{\Large{N}}&$\sim10^{18}\div10^{21}$&$1\div10$\,nm&$\sim 1$&$\sim 1$&$\sim 1\div 3$&$\sim10^{14}\div 10^{15}$&$4.8\div5.2$& \\ 
  \hline\hline 
  \end{tabular} 
  \caption{Summary table of the values of the fitting
    parameters obtained from the simulation
    $I$-$V$ characteristics
    for $n$- and $p$-\ce{Si}/\ce{ZrO2}/\ce{Ni} structures
    in different models: (F) Frenkel model (\ref{eq:P:Frenkel}), (Hill)
    Hill model (Pool law) (\ref{eq:P:Hill}),
    (MPTI) multiphonon trap ionization (\ref{e:P:Makram}),
    (PAT) phonon-assisted tunneling between traps
    (\ref{eq:P:Nasyrov}). The last column represent
    ranges of expected (reasonable) values (from calculation or
    literature) if any.}
 \label{t:values}
\end{table*}

Phonon-assisted tunneling between traps adequately
describes charge transport in \ce{Hf_{0.5}Zr_{0.5}O2} films
on $n$-\ce{Si} and $p$-\ce{Si} substrates.
Taking these into account,
we conclude that the model
of phonon-assisted tunneling between traps
describe charge transport in \ce{Hf_{0.5}Zr_{0.5}O2} films.
Energy parameters of traps, such as
the thermal trap energy of $1.25$\,eV
and the optical trap energy of $2.5$\,eV,
are similar to that in
binary oxides
\ce{HfO2} \cite{Perevalov:APL104:071904, HfO2:Tranport:2014},
and \ce{ZrO2} \cite{Perevalov:ZrO2:2014}.

\begin{figure}
  \includegraphics[width=\columnwidth]{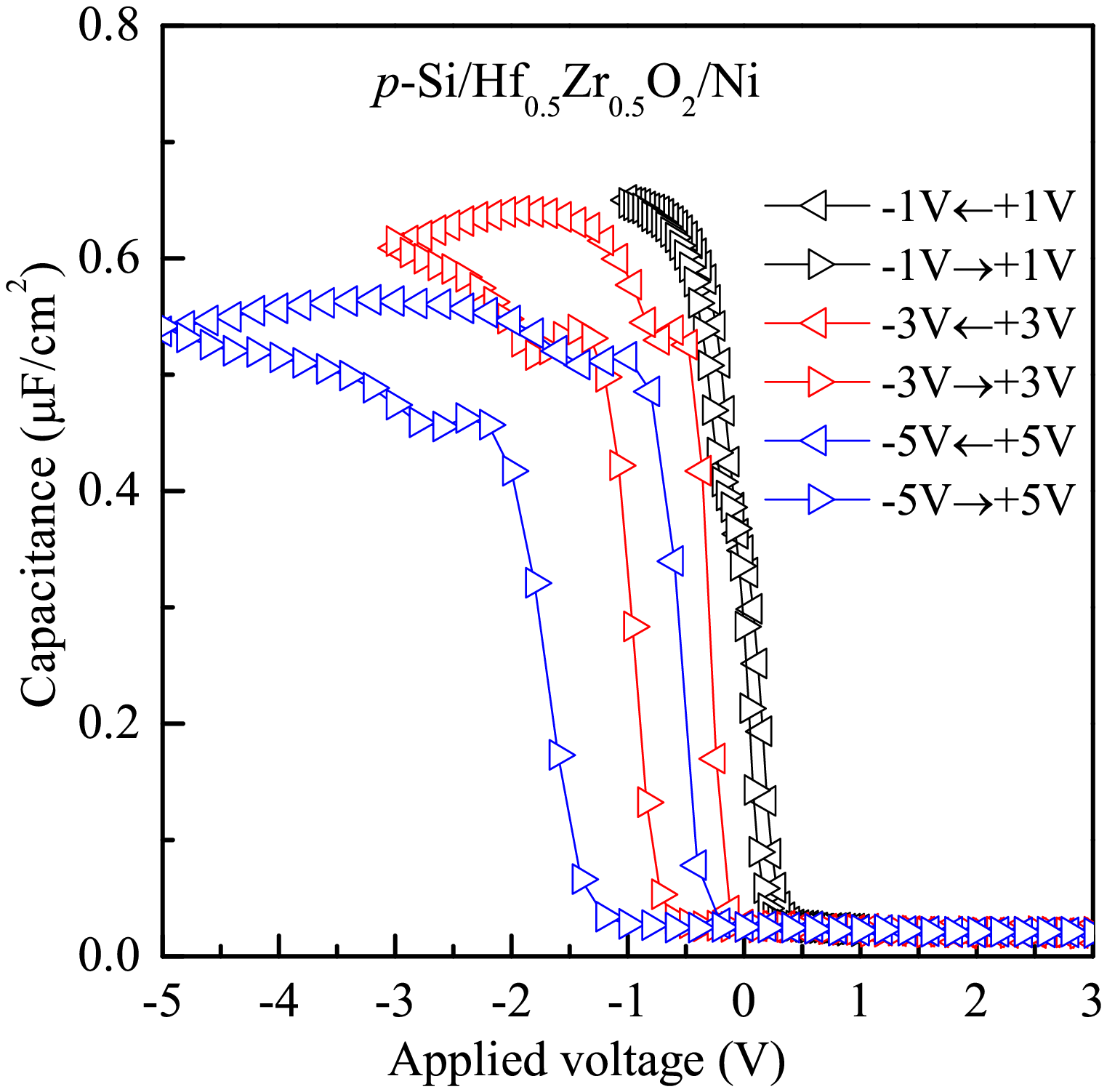}
  \caption{Experimental ca\-pa\-ci\-tance-vol\-tage
    characteristics
    in $p$-\ce{Si}/\ce{Hf_{0.5}Zr_{0.5}O2}/\ce{Ni} structures
    at different voltage limits.}
  \label{f:p-Si_HfZrO-20nm_Ni:CV}
\end{figure}

Ca\-pa\-ci\-tance-vol\-tage
($C$-$V$) measurements (Fig.~\ref{f:p-Si_HfZrO-20nm_Ni:CV})
show that increasing of voltage amplitude leads to
shift of the hysteresis to negative voltages.
This phenomenon might be caused by
holes trapping on \ce{Hf_{0.5}Zr_{0.5}O2} interfaces or
in the bulk of the dielectric.
$C$-$V$ shift allows us to valuate the density
of the filled hole traps as
$n_mathrm{h}^\mathrm{b}\lesssim 3\times 10^{18}$\,cm$^{-3}$
in the bulk or
$n_mathrm{h}^\mathrm{s}\lesssim 6\times 10^{12}$\,cm$^{-2}$
on the surface states. It is not possible to separate
percentage of the charge on the surface and in the bulk,
unusual $C$-$V$ behavior of reducing the maximum
capacity with an increase in the voltage amplitude
indicates that the the surface charge is
significant.
However, it should be noted that
the number of filled bulk traps 
$n_mathrm{h}^\mathrm{b}\lesssim 3\times 10^{18}$\,cm$^{-3}$
is much lower than total trap density of
$N=3\times 10^{19}$\,cm$^{-3}$.
The possible explanation of this difference
is the Coulomb repulsion of the charged particles
with forming of Wigner-glass-like
structures \cite{high-k:Wigner:apl96:263510}
in the bulk of \ce{Hf_{0.5}Zr_{0.5}O2}.

It was shown that oxygen vacancies
are responsible for charge transport via
\ce{HfO2} and \ce{ZrO2} \cite{Perevalov:APL104:071904, HfO2:Tranport:2014, Perevalov:ZrO2:2014}.
Thus, one can expect that oxygen vacancies
are responsible for charge transport
in \ce{Hf_{0.5}Zr_{0.5}O2} too.
To confirm this hypothesis experiments on
photoluminescence 
and quantum-chemical simulations are required.

Recently, it was reported that
orthorhombic crystalline phase of
high-$\kappa$ \ce{Hf_{0.5}Zr_{0.5}O2}
thin films can be ferroelectrics,
being perspective material for
application to ferroelectric random access memory
(FeRAM) \cite{FeRAM:HfZrO:apl99:112901, FeRAM:HfZrO:apl102:112914, FeRAM:HfZrO:apl102:242905}.
Despite that FeRAM has many advantages,
retention characteristics of FeRAM devices
much to be desired because
of the depolarization effect \cite{FeRAM:HfZrO:edl35:138}.
A possible reason of the depolarization effect
is charge leakage via traps of the dielectric.
To confirm or refute this hypothesis,
transport properties of \ce{Hf_{0.5}Zr_{0.5}O2}
in amorphous and ferroelectric phases
must be compared.

To summarize, we examined the transport mechanisms of
amorphous solid solution \ce{Hf_{0.5}Zr_{0.5}O2}.
It was demonstrated that all charge transport models
such as Frenkel model, Hill model,
multiphonon trap ionization, and
phonon-assisted tunneling between traps
describe experiment data formally, qualitatively,
while only
phonon-assisted tunneling between traps
describes the charge transport in \ce{Hf_{0.5}Zr_{0.5}O2}
quantitatively.
Comparing experimental current-voltage characteristics
with results of simulations revealed energy
parameters of the charge traps in
\ce{Hf_{0.5}Zr_{0.5}O2}:
the thermal trap energy of $1.25$\,eV
and the optical trap energy of $2.5$\,eV,
that are equal to that for 
\ce{HfO2} and \ce{ZrO2}.

This work was particularly supported by
Ministry of Science and Technology, Taiwan
(grant No.~NSC103-2923-E-009-002-MY3)
(growing test structures, preparing samples, performing transport measurements),
and by the Russian Science Foundation (grant No.~14-19-00192)
(calculations, modeling).

\bibliographystyle{apsrev4-1}
\bibliography{IEEEabrv,../../../bibtex/Technique,../../../bibtex/GeO2,../../../bibtex/TiO2,../../../bibtex/HfO2,../../../bibtex/SiO2,../../../bibtex/TaOx,../../../bibtex/Theory,../../../bibtex/percollation,../../../bibtex/Memristor,../../../bibtex/computing,../../../bibtex/ZrO2,../../../bibtex/Al2O3,../../../bibtex/minecite,../../../bibtex/FeRAM,../../../bibtex/high-k,../../../bibtex/Si3N4}

 \providecommand{\noop}[1]{}  \providecommand{\noop}[1]{}
  \providecommand{\noop}[1]{}  \providecommand{\noop}[1]{}
\begin{thebibliography}{19}%
\makeatletter
\providecommand \@ifxundefined [1]{%
 \@ifx{#1\undefined}
}%
\providecommand \@ifnum [1]{%
 \ifnum #1\expandafter \@firstoftwo
 \else \expandafter \@secondoftwo
 \fi
}%
\providecommand \@ifx [1]{%
 \ifx #1\expandafter \@firstoftwo
 \else \expandafter \@secondoftwo
 \fi
}%
\providecommand \natexlab [1]{#1}%
\providecommand \enquote  [1]{``#1''}%
\providecommand \bibnamefont  [1]{#1}%
\providecommand \bibfnamefont [1]{#1}%
\providecommand \citenamefont [1]{#1}%
\providecommand \href@noop [0]{\@secondoftwo}%
\providecommand \href [0]{\begingroup \@sanitize@url \@href}%
\providecommand \@href[1]{\@@startlink{#1}\@@href}%
\providecommand \@@href[1]{\endgroup#1\@@endlink}%
\providecommand \@sanitize@url [0]{\catcode `\\12\catcode `\$12\catcode
  `\&12\catcode `\#12\catcode `\^12\catcode `\_12\catcode `\%12\relax}%
\providecommand \@@startlink[1]{}%
\providecommand \@@endlink[0]{}%
\providecommand \url  [0]{\begingroup\@sanitize@url \@url }%
\providecommand \@url [1]{\endgroup\@href {#1}{\urlprefix }}%
\providecommand \urlprefix  [0]{URL }%
\providecommand \Eprint [0]{\href }%
\providecommand \doibase [0]{http://dx.doi.org/}%
\providecommand \selectlanguage [0]{\@gobble}%
\providecommand \bibinfo  [0]{\@secondoftwo}%
\providecommand \bibfield  [0]{\@secondoftwo}%
\providecommand \translation [1]{[#1]}%
\providecommand \BibitemOpen [0]{}%
\providecommand \bibitemStop [0]{}%
\providecommand \bibitemNoStop [0]{.\EOS\space}%
\providecommand \EOS [0]{\spacefactor3000\relax}%
\providecommand \BibitemShut  [1]{\csname bibitem#1\endcsname}%
\let\auto@bib@innerbib\@empty
\bibitem [{\citenamefont {Novikov}(2013)}]{NovikovHfO2:JAP113:024109}%
  \BibitemOpen
  \bibfield  {author} {\bibinfo {author} {\bibfnamefont {Y.~N.}\ \bibnamefont
  {Novikov}},\ }\href {\doibase 10.1063/1.4775407} {\bibfield  {journal}
  {\bibinfo  {journal} {Journal of Applied Physics}\ }\textbf {\bibinfo
  {volume} {113}},\ \bibinfo {eid} {024109} (\bibinfo {year}
  {2013})}\BibitemShut {NoStop}%
\bibitem [{\citenamefont {Islamov}\ \emph {et~al.}(2014)\citenamefont
  {Islamov}, \citenamefont {Gritsenko}, \citenamefont {Cheng},\ and\
  \citenamefont {Chin}}]{HfO2:Tranport:2014}%
  \BibitemOpen
  \bibfield  {author} {\bibinfo {author} {\bibfnamefont {D.~R.}\ \bibnamefont
  {Islamov}}, \bibinfo {author} {\bibfnamefont {V.~A.}\ \bibnamefont
  {Gritsenko}}, \bibinfo {author} {\bibfnamefont {C.~H.}\ \bibnamefont
  {Cheng}}, \ and\ \bibinfo {author} {\bibfnamefont {A.}~\bibnamefont {Chin}},\
  }\href {\doibase 10.1063/1.4903169} {\bibfield  {journal} {\bibinfo
  {journal} {Applied Physics Letters}\ }\textbf {\bibinfo {volume} {105}},\
  \bibinfo {pages} {222901} (\bibinfo {year} {2014})},\ \Eprint
  {http://arxiv.org/abs/1409.6887} {arXiv:1409.6887 [cond-mat.mtrl-sci]}
  \BibitemShut {NoStop}%
\bibitem [{\citenamefont {Jegert}\ \emph {et~al.}(2011)\citenamefont {Jegert},
  \citenamefont {Kersch}, \citenamefont {Weinreich},\ and\ \citenamefont
  {Lugli}}]{ZrO2:Transport:jap109:014504}%
  \BibitemOpen
  \bibfield  {author} {\bibinfo {author} {\bibfnamefont {G.}~\bibnamefont
  {Jegert}}, \bibinfo {author} {\bibfnamefont {A.}~\bibnamefont {Kersch}},
  \bibinfo {author} {\bibfnamefont {W.}~\bibnamefont {Weinreich}}, \ and\
  \bibinfo {author} {\bibfnamefont {P.}~\bibnamefont {Lugli}},\ }\href
  {\doibase 10.1063/1.3531538} {\bibfield  {journal} {\bibinfo  {journal}
  {Journal of Applied Physics}\ }\textbf {\bibinfo {volume} {109}},\ \bibinfo
  {eid} {014504} (\bibinfo {year} {2011}),\ 10.1063/1.3531538}\BibitemShut
  {NoStop}%
\bibitem [{\citenamefont {Islamov}\ \emph {et~al.}(shed)\citenamefont
  {Islamov}, \citenamefont {Gritsenko}, \citenamefont {Cheng},\ and\
  \citenamefont {Chin}}]{ZrO2:transport:2015}%
  \BibitemOpen
  \bibfield  {author} {\bibinfo {author} {\bibfnamefont {D.~R.}\ \bibnamefont
  {Islamov}}, \bibinfo {author} {\bibfnamefont {V.~A.}\ \bibnamefont
  {Gritsenko}}, \bibinfo {author} {\bibfnamefont {C.~H.}\ \bibnamefont
  {Cheng}}, \ and\ \bibinfo {author} {\bibfnamefont {A.}~\bibnamefont {Chin}},\
  }\href@noop {} {\  (\bibinfo {year} {\noop{2015}unpublished})}\BibitemShut
  {NoStop}%
\bibitem [{\citenamefont {Giannozzi}\ \emph {et~al.}(2009)\citenamefont
  {Giannozzi}, \citenamefont {Baroni}, \citenamefont {Bonini}, \citenamefont
  {Calandra}, \citenamefont {Car}, \citenamefont {Cavazzoni}, \citenamefont
  {Ceresoli}, \citenamefont {Chiarotti}, \citenamefont {Cococcioni},
  \citenamefont {Dabo}, \citenamefont {{Dal Corso}}, \citenamefont {{de
  Gironcoli}}, \citenamefont {Fabris}, \citenamefont {Fratesi}, \citenamefont
  {Gebauer}, \citenamefont {Gerstmann}, \citenamefont {Gougoussis},
  \citenamefont {Kokalj}, \citenamefont {Lazzeri}, \citenamefont
  {Martin-Samos}, \citenamefont {Marzari}, \citenamefont {Mauri}, \citenamefont
  {Mazzarello}, \citenamefont {Paolini}, \citenamefont {Pasquarello},
  \citenamefont {Paulatto}, \citenamefont {Sbraccia}, \citenamefont {Scandolo},
  \citenamefont {Sclauzero}, \citenamefont {Seitsonen}, \citenamefont
  {Smogunov}, \citenamefont {Umari},\ and\ \citenamefont
  {Wentzcovitch}}]{QE-2009}%
  \BibitemOpen
  \bibfield  {author} {\bibinfo {author} {\bibfnamefont {P.}~\bibnamefont
  {Giannozzi}}, \bibinfo {author} {\bibfnamefont {S.}~\bibnamefont {Baroni}},
  \bibinfo {author} {\bibfnamefont {N.}~\bibnamefont {Bonini}}, \bibinfo
  {author} {\bibfnamefont {M.}~\bibnamefont {Calandra}}, \bibinfo {author}
  {\bibfnamefont {R.}~\bibnamefont {Car}}, \bibinfo {author} {\bibfnamefont
  {C.}~\bibnamefont {Cavazzoni}}, \bibinfo {author} {\bibfnamefont
  {D.}~\bibnamefont {Ceresoli}}, \bibinfo {author} {\bibfnamefont {G.~L.}\
  \bibnamefont {Chiarotti}}, \bibinfo {author} {\bibfnamefont {M.}~\bibnamefont
  {Cococcioni}}, \bibinfo {author} {\bibfnamefont {I.}~\bibnamefont {Dabo}},
  \bibinfo {author} {\bibfnamefont {A.}~\bibnamefont {{Dal Corso}}}, \bibinfo
  {author} {\bibfnamefont {S.}~\bibnamefont {{de Gironcoli}}}, \bibinfo
  {author} {\bibfnamefont {S.}~\bibnamefont {Fabris}}, \bibinfo {author}
  {\bibfnamefont {G.}~\bibnamefont {Fratesi}}, \bibinfo {author} {\bibfnamefont
  {R.}~\bibnamefont {Gebauer}}, \bibinfo {author} {\bibfnamefont
  {U.}~\bibnamefont {Gerstmann}}, \bibinfo {author} {\bibfnamefont
  {C.}~\bibnamefont {Gougoussis}}, \bibinfo {author} {\bibfnamefont
  {A.}~\bibnamefont {Kokalj}}, \bibinfo {author} {\bibfnamefont
  {M.}~\bibnamefont {Lazzeri}}, \bibinfo {author} {\bibfnamefont
  {L.}~\bibnamefont {Martin-Samos}}, \bibinfo {author} {\bibfnamefont
  {N.}~\bibnamefont {Marzari}}, \bibinfo {author} {\bibfnamefont
  {F.}~\bibnamefont {Mauri}}, \bibinfo {author} {\bibfnamefont
  {R.}~\bibnamefont {Mazzarello}}, \bibinfo {author} {\bibfnamefont
  {S.}~\bibnamefont {Paolini}}, \bibinfo {author} {\bibfnamefont
  {A.}~\bibnamefont {Pasquarello}}, \bibinfo {author} {\bibfnamefont
  {L.}~\bibnamefont {Paulatto}}, \bibinfo {author} {\bibfnamefont
  {C.}~\bibnamefont {Sbraccia}}, \bibinfo {author} {\bibfnamefont
  {S.}~\bibnamefont {Scandolo}}, \bibinfo {author} {\bibfnamefont
  {G.}~\bibnamefont {Sclauzero}}, \bibinfo {author} {\bibfnamefont {A.~P.}\
  \bibnamefont {Seitsonen}}, \bibinfo {author} {\bibfnamefont {A.}~\bibnamefont
  {Smogunov}}, \bibinfo {author} {\bibfnamefont {P.}~\bibnamefont {Umari}}, \
  and\ \bibinfo {author} {\bibfnamefont {R.~M.}\ \bibnamefont {Wentzcovitch}},\
  }\href {\doibase 10.1088/0953-8984/21/39/395502} {\bibfield  {journal}
  {\bibinfo  {journal} {Journal of Physics: Condensed Matter}\ }\textbf
  {\bibinfo {volume} {21}},\ \bibinfo {pages} {395502} (\bibinfo {year}
  {2009})}\BibitemShut {NoStop}%
\bibitem [{\citenamefont
  {Dutta}(2009)}]{ZrO2:1principle:epsilon:jap105:106103}%
  \BibitemOpen
  \bibfield  {author} {\bibinfo {author} {\bibfnamefont {G.}~\bibnamefont
  {Dutta}},\ }\href {\doibase 10.1063/1.3117829} {\bibfield  {journal}
  {\bibinfo  {journal} {Journal of Applied Physics}\ }\textbf {\bibinfo
  {volume} {105}},\ \bibinfo {eid} {106103} (\bibinfo {year} {2009}),\
  10.1063/1.3117829}\BibitemShut {NoStop}%
\bibitem [{\citenamefont {Frenkel}(1938)}]{PhysRev:54:647}%
  \BibitemOpen
  \bibfield  {author} {\bibinfo {author} {\bibfnamefont {J.}~\bibnamefont
  {Frenkel}},\ }\href {\doibase 10.1103/PhysRev.54.647} {\bibfield  {journal}
  {\bibinfo  {journal} {Physical Review}\ }\textbf {\bibinfo {volume} {54}},\
  \bibinfo {pages} {647} (\bibinfo {year} {1938})}\BibitemShut {NoStop}%
\bibitem [{\citenamefont {Kar}(2013)}]{high-k:Samares:2013}%
  \BibitemOpen
  \bibinfo {editor} {\bibfnamefont {S.}~\bibnamefont {Kar}},\ ed.,\ \href@noop
  {} {{\selectlanguage {english}\emph {\bibinfo {title} {High Permittivity Gate
  Dielectric Materials}}}},\ \bibinfo {series} {Springer Series in Advanced
  Microelectronics}, Vol.~\bibinfo {volume} {43}\ (\bibinfo  {publisher}
  {Springer},\ \bibinfo {year} {2013})\BibitemShut {NoStop}%
\bibitem [{\citenamefont {Rignanese}\ \emph {et~al.}(2001)\citenamefont
  {Rignanese}, \citenamefont {Detraux}, \citenamefont {Gonze},\ and\
  \citenamefont {Pasquarello}}]{ZrO2:abinitio:PRB64:134301}%
  \BibitemOpen
  \bibfield  {author} {\bibinfo {author} {\bibfnamefont {G.-M.}\ \bibnamefont
  {Rignanese}}, \bibinfo {author} {\bibfnamefont {F.}~\bibnamefont {Detraux}},
  \bibinfo {author} {\bibfnamefont {X.}~\bibnamefont {Gonze}}, \ and\ \bibinfo
  {author} {\bibfnamefont {A.}~\bibnamefont {Pasquarello}},\ }\href {\doibase
  10.1103/PhysRevB.64.134301} {\bibfield  {journal} {\bibinfo  {journal}
  {Physical Review B}\ }\textbf {\bibinfo {volume} {64}},\ \bibinfo {pages}
  {134301} (\bibinfo {year} {2001})}\BibitemShut {NoStop}%
\bibitem [{\citenamefont {Hill}(1971)}]{PhilMag:23:59}%
  \BibitemOpen
  \bibfield  {author} {\bibinfo {author} {\bibfnamefont {R.~M.}\ \bibnamefont
  {Hill}},\ }\href {\doibase 10.1080/14786437108216365} {\bibfield  {journal}
  {\bibinfo  {journal} {Philosophical Magazine}\ }\textbf {\bibinfo {volume}
  {23}},\ \bibinfo {pages} {59} (\bibinfo {year} {1971})}\BibitemShut {NoStop}%
\bibitem [{\citenamefont {Makram-Ebeid}\ and\ \citenamefont
  {Lannoo}(1982)}]{PhysRevB:25:6406}%
  \BibitemOpen
  \bibfield  {author} {\bibinfo {author} {\bibfnamefont {S.}~\bibnamefont
  {Makram-Ebeid}}\ and\ \bibinfo {author} {\bibfnamefont {M.}~\bibnamefont
  {Lannoo}},\ }\href {\doibase 10.1103/PhysRevB.25.6406} {\bibfield  {journal}
  {\bibinfo  {journal} {Physical Review B}\ }\textbf {\bibinfo {volume} {25}},\
  \bibinfo {pages} {6406} (\bibinfo {year} {1982})}\BibitemShut {NoStop}%
\bibitem [{\citenamefont {Nasyrov}\ and\ \citenamefont
  {Gritsenko}(2011)}]{JAP:109:093705}%
  \BibitemOpen
  \bibfield  {author} {\bibinfo {author} {\bibfnamefont {K.~A.}\ \bibnamefont
  {Nasyrov}}\ and\ \bibinfo {author} {\bibfnamefont {V.~A.}\ \bibnamefont
  {Gritsenko}},\ }\href {\doibase 10.1063/1.3587452} {\bibfield  {journal}
  {\bibinfo  {journal} {Journal of Applied Physics}\ }\textbf {\bibinfo
  {volume} {109}},\ \bibinfo {pages} {093705} (\bibinfo {year}
  {2011})}\BibitemShut {NoStop}%
\bibitem [{\citenamefont {Perevalov}\ \emph
  {et~al.}(2014{\natexlab{a}})\citenamefont {Perevalov}, \citenamefont {Aliev},
  \citenamefont {Gritsenko}, \citenamefont {Saraev}, \citenamefont {Kaichev},
  \citenamefont {Ivanova},\ and\ \citenamefont
  {Zamoryanskaya}}]{Perevalov:APL104:071904}%
  \BibitemOpen
  \bibfield  {author} {\bibinfo {author} {\bibfnamefont {T.~V.}\ \bibnamefont
  {Perevalov}}, \bibinfo {author} {\bibfnamefont {V.~S.}\ \bibnamefont
  {Aliev}}, \bibinfo {author} {\bibfnamefont {V.~A.}\ \bibnamefont
  {Gritsenko}}, \bibinfo {author} {\bibfnamefont {A.~A.}\ \bibnamefont
  {Saraev}}, \bibinfo {author} {\bibfnamefont {V.~V.}\ \bibnamefont {Kaichev}},
  \bibinfo {author} {\bibfnamefont {E.~V.}\ \bibnamefont {Ivanova}}, \ and\
  \bibinfo {author} {\bibfnamefont {M.~V.}\ \bibnamefont {Zamoryanskaya}},\
  }\href {\doibase 10.1063/1.4865259} {\bibfield  {journal} {\bibinfo
  {journal} {Applied Physics Letters}\ }\textbf {\bibinfo {volume} {104}},\
  \bibinfo {eid} {071904} (\bibinfo {year} {2014}{\natexlab{a}}),\
  10.1063/1.4865259}\BibitemShut {NoStop}%
\bibitem [{\citenamefont {Perevalov}\ \emph
  {et~al.}(2014{\natexlab{b}})\citenamefont {Perevalov}, \citenamefont
  {Gulyaev}, \citenamefont {Aliev}, \citenamefont {Zhuravlev}, \citenamefont
  {Gritsenko},\ and\ \citenamefont {Yelisseyev}}]{Perevalov:ZrO2:2014}%
  \BibitemOpen
  \bibfield  {author} {\bibinfo {author} {\bibfnamefont {T.~V.}\ \bibnamefont
  {Perevalov}}, \bibinfo {author} {\bibfnamefont {D.~V.}\ \bibnamefont
  {Gulyaev}}, \bibinfo {author} {\bibfnamefont {V.~S.}\ \bibnamefont {Aliev}},
  \bibinfo {author} {\bibfnamefont {K.~S.}\ \bibnamefont {Zhuravlev}}, \bibinfo
  {author} {\bibfnamefont {V.~A.}\ \bibnamefont {Gritsenko}}, \ and\ \bibinfo
  {author} {\bibfnamefont {A.~P.}\ \bibnamefont {Yelisseyev}},\ }\href
  {\doibase 10.1063/1.4905105} {\bibfield  {journal} {\bibinfo  {journal}
  {Journal of Applied Physics}\ }\textbf {\bibinfo {volume} {116}},\ \bibinfo
  {pages} {244109} (\bibinfo {year} {2014}{\natexlab{b}})}\BibitemShut
  {NoStop}%
\bibitem [{\citenamefont {Shaimeev}\ \emph {et~al.}(2010)\citenamefont
  {Shaimeev}, \citenamefont {Gritsenko},\ and\ \citenamefont
  {Wong}}]{high-k:Wigner:apl96:263510}%
  \BibitemOpen
  \bibfield  {author} {\bibinfo {author} {\bibfnamefont {S.~S.}\ \bibnamefont
  {Shaimeev}}, \bibinfo {author} {\bibfnamefont {V.~A.}\ \bibnamefont
  {Gritsenko}}, \ and\ \bibinfo {author} {\bibfnamefont {H.}~\bibnamefont
  {Wong}},\ }\href {\doibase 10.1063/1.3458832} {\bibfield  {journal} {\bibinfo
   {journal} {Applied Physics Letters}\ }\textbf {\bibinfo {volume} {96}},\
  \bibinfo {eid} {263510} (\bibinfo {year} {2010}),\
  10.1063/1.3458832}\BibitemShut {NoStop}%
\bibitem [{\citenamefont {M\"{u}ller}\ \emph {et~al.}(2011)\citenamefont
  {M\"{u}ller}, \citenamefont {B\"{o}scke}, \citenamefont {Br\"{a}uhaus},
  \citenamefont {Schr\"{o}der}, \citenamefont {B\"{o}ttger}, \citenamefont
  {Sundqvist}, \citenamefont {K\"{u}cher}, \citenamefont {Mikolajick},\ and\
  \citenamefont {Frey}}]{FeRAM:HfZrO:apl99:112901}%
  \BibitemOpen
  \bibfield  {author} {\bibinfo {author} {\bibfnamefont {J.}~\bibnamefont
  {M\"{u}ller}}, \bibinfo {author} {\bibfnamefont {T.~S.}\ \bibnamefont
  {B\"{o}scke}}, \bibinfo {author} {\bibfnamefont {D.}~\bibnamefont
  {Br\"{a}uhaus}}, \bibinfo {author} {\bibfnamefont {U.}~\bibnamefont
  {Schr\"{o}der}}, \bibinfo {author} {\bibfnamefont {U.}~\bibnamefont
  {B\"{o}ttger}}, \bibinfo {author} {\bibfnamefont {J.}~\bibnamefont
  {Sundqvist}}, \bibinfo {author} {\bibfnamefont {P.}~\bibnamefont
  {K\"{u}cher}}, \bibinfo {author} {\bibfnamefont {T.}~\bibnamefont
  {Mikolajick}}, \ and\ \bibinfo {author} {\bibfnamefont {L.}~\bibnamefont
  {Frey}},\ }\href {\doibase 10.1063/1.3636417} {\bibfield  {journal} {\bibinfo
   {journal} {Applied Physics Letters}\ }\textbf {\bibinfo {volume} {99}},\
  \bibinfo {pages} {112901} (\bibinfo {year} {2011})}\BibitemShut {NoStop}%
\bibitem [{\citenamefont {Hyuk~Park}\ \emph
  {et~al.}(2013{\natexlab{a}})\citenamefont {Hyuk~Park}, \citenamefont
  {Joon~Kim}, \citenamefont {Jin~Kim}, \citenamefont {Lee}, \citenamefont
  {Kyeom~Kim},\ and\ \citenamefont {Seong~Hwang}}]{FeRAM:HfZrO:apl102:112914}%
  \BibitemOpen
  \bibfield  {author} {\bibinfo {author} {\bibfnamefont {M.}~\bibnamefont
  {Hyuk~Park}}, \bibinfo {author} {\bibfnamefont {H.}~\bibnamefont {Joon~Kim}},
  \bibinfo {author} {\bibfnamefont {Y.}~\bibnamefont {Jin~Kim}}, \bibinfo
  {author} {\bibfnamefont {W.}~\bibnamefont {Lee}}, \bibinfo {author}
  {\bibfnamefont {H.}~\bibnamefont {Kyeom~Kim}}, \ and\ \bibinfo {author}
  {\bibfnamefont {C.}~\bibnamefont {Seong~Hwang}},\ }\href {\doibase
  10.1063/1.4798265} {\bibfield  {journal} {\bibinfo  {journal} {Applied
  Physics Letters}\ }\textbf {\bibinfo {volume} {102}},\ \bibinfo {pages}
  {112914} (\bibinfo {year} {2013}{\natexlab{a}})}\BibitemShut {NoStop}%
\bibitem [{\citenamefont {Hyuk~Park}\ \emph
  {et~al.}(2013{\natexlab{b}})\citenamefont {Hyuk~Park}, \citenamefont
  {Joon~Kim}, \citenamefont {Jin~Kim}, \citenamefont {Lee}, \citenamefont
  {Moon},\ and\ \citenamefont {Seong~Hwang}}]{FeRAM:HfZrO:apl102:242905}%
  \BibitemOpen
  \bibfield  {author} {\bibinfo {author} {\bibfnamefont {M.}~\bibnamefont
  {Hyuk~Park}}, \bibinfo {author} {\bibfnamefont {H.}~\bibnamefont {Joon~Kim}},
  \bibinfo {author} {\bibfnamefont {Y.}~\bibnamefont {Jin~Kim}}, \bibinfo
  {author} {\bibfnamefont {W.}~\bibnamefont {Lee}}, \bibinfo {author}
  {\bibfnamefont {T.}~\bibnamefont {Moon}}, \ and\ \bibinfo {author}
  {\bibfnamefont {C.}~\bibnamefont {Seong~Hwang}},\ }\href {\doibase
  10.1063/1.4811483} {\bibfield  {journal} {\bibinfo  {journal} {Applied
  Physics Letters}\ }\textbf {\bibinfo {volume} {102}},\ \bibinfo {pages}
  {242905} (\bibinfo {year} {2013}{\natexlab{b}})}\BibitemShut {NoStop}%
\bibitem [{\citenamefont {Cheng}\ and\ \citenamefont
  {Chin}(2014)}]{FeRAM:HfZrO:edl35:138}%
  \BibitemOpen
  \bibfield  {author} {\bibinfo {author} {\bibfnamefont {C.-H.}\ \bibnamefont
  {Cheng}}\ and\ \bibinfo {author} {\bibfnamefont {A.}~\bibnamefont {Chin}},\
  }\href {\doibase 10.1109/LED.2013.2290117} {\bibfield  {journal} {\bibinfo
  {journal} {{IEEE} Electron Device Lett.}\ }\textbf {\bibinfo {volume} {35}},\
  \bibinfo {pages} {138} (\bibinfo {year} {2014})}\BibitemShut {NoStop}%
\end{thebibliography}%

\end{document}